\documentclass[a4paper,11pt]{article}
\usepackage{pos}

\title{Data Quality Monitoring system of the Baikal-GVD experiment}
 \ShortTitle{DQM in the Baikal-GVD}

\author[a]{V.A.~Allakhverdyan}
\author[b]{A.D.~Avrorin}
\author[b]{A.V.~Avrorin}
\author[b]{V.M.~Aynutdinov}
\author[c]{R.~Bannasch}
\author[d]{Z.~Barda\v{c}ov\'{a}}
\author[a]{I.A.~Belolaptikov}
\author[a]{I.V.~Borina}
\author[a,1]{V.B.~Brudanin}
\author[e]{N.M.~Budnev}
\author[a]{V.Y.~Dik}
\author[b]{G.V.~Domogatsky}
\author[b]{A.A.~Doroshenko}
\author[a,d]{R.~Dvornick\'{y}}
\author[e]{A.N.~Dyachok}
\author[b]{Zh.-A.M.~Dzhilkibaev}
\author[d]{E.~Eckerov\'{a}}
\author[a]{T.V.~Elzhov}
\author[f]{L.~Fajt}
\author[g,1]{S.V.~Fialkovski}
\author[e]{A.R.~Gafarov}
\author[b]{K.V.~Golubkov}
\author[a]{N.S.~Gorshkov}
\author[e]{T.I.~Gress}
\author[a]{M.S.~Katulin}
\author[c]{K.G.~Kebkal}
\author[c]{O.G.~Kebkal}
\author[a]{E.V.~Khramov}
\author[a]{M.M.~Kolbin}
\author[a]{K.V.~Konischev}
\author[h]{K.A.~Kopa\'{n}ski}
\author[a]{A.V.~Korobchenko}
\author[b]{A.P.~Koshechkin}
\author[i]{V.A.~Kozhin}
\author[a]{M.V.~Kruglov}
\author[b]{M.K.~Kryukov}
\author[g]{V.F.~Kulepov}
\author[h]{Pa.~Malecki}
\author[a]{Y.M.~Malyshkin}
\author[b]{M.B.~Milenin}
\author[e]{R.R.~Mirgazov}
\author[a]{D.V.~Naumov}
\author[a]{V.~Nazari}
\author[h]{W.~Noga}
\author[b]{D.P.~Petukhov}
\author[a]{E.N.~Pliskovsky}
\author[j]{M.I.~Rozanov}
\author[a]{V.D.~Rushay}
\author[e]{E.V.~Ryabov}
\author[b]{G.B.~Safronov}
\author[a]{B.A.~Shaybonov}
\author[b]{M.D.~Shelepov}
\author[a,d,f]{F.~\v{S}imkovic}
\author[a]{A.E.~Sirenko}
\author[i]{A.V.~Skurikhin}
\author[a]{A.G.~Solovjev}
\author*[a]{M.N.~Sorokovikov}
\author[f]{I.~\v{S}tekl}
\author[b]{A.P.~Stromakov}
\author[a]{E.O.~Sushenok}
\author[b]{O.V.~Suvorova}
\author[e]{V.A.~Tabolenko}
\author[e]{B.A.~Tarashansky}
\author[a]{Y.V.~Yablokova}
\author[c]{S.A.~Yakovlev}
\author[b]{D.N.~Zaborov}

\affiliation[a]{Joint Institute for Nuclear Research, Dubna, Russia}
\affiliation[b]{Institute for Nuclear Research, Russian Academy of Sciences, Moscow, Russia}
\affiliation[c]{EvoLogics GmbH, Berlin, Germany}
\affiliation[d]{Comenius University, Bratislava, Slovakia}
\affiliation[e]{Irkutsk State University, Irkutsk, Russia}
\affiliation[f]{Czech Technical University in Prague, Prague, Czech Republic}
\affiliation[g]{Nizhny Novgorod State Technical University, Nizhny Novgorod, Russia}
\affiliation[h]{Institute of Nuclear Physics of Polish Academy of Sciences (IFJ~PAN), Krak\'{o}w, Poland}
\affiliation[i]{Skobeltsyn Institute of Nuclear Physics, Moscow State University, Moscow, Russia}
\affiliation[j]{St.~Petersburg State Marine Technical University, St.Petersburg, Russia}

\note{Deceased.}

\emailAdd{sorokovikov@jinr.ru}

\abstract{The main purpose of the Baikal-GVD Data Quality Monitoring (DQM) system is to monitor the status of the detector and collected data. The system estimates quality of the recorded signals and performs the data validation. The DQM system is integrated with the Baikal-GVD’s unified software framework ("BARS") and operates in quasi-online manner. This allows us to react promptly and effectively to the changes in the telescope conditions.}

\FullConference{37$^{\rm{th}}$ International Cosmic Ray Conference (ICRC 2021)\\
		July 12th -- 23rd, 2021\\
		Online -- Berlin, Germany}

\begin{document}
\maketitle

\section{Introduction}
Baikal-GVD is a km$^3$-scale neutrino telescope currently under construction in Lake Baikal~\cite{GVD1, GVD2}. It produces about 100GB of data every day, and verification of large amount of data is necessary. The main purpose of the Baikal-GVD Data Quality Monitoring (DQM) system is to monitor the status of the detector and collected data. The system estimates quality of the recorded signals and performs the data validation. The DQM system is integrated with the Baikal-GVD’s unified software framework "BARS"~\cite{BARS} and operates in quasi-online manner. The characteristics of data records under monitoring form two groups of parameters related to the distributions described the Poissonian character of events flow, and also the signal charge measurements. The first group consists of the exponential distribution, uniform distribution and Poissonian one. The second group comprises the single photoelectron (p.e) distribution, channel noise rate, and trigger thresholds. We fit the distributions with expected functions by means of the minimum chi square method, and $\chi^2/NDF$ parameter is used to obtain quality estimations.

\section{Estimation of the Poissonian character of events flow}
The signals generated by atmospheric muons and lake random noise are dominated in the total data flow of the telescope. Detection of this type of events is Poissonian-like process and recorded telescope data obey the next three distribution – exponential, Poissonian, and uniform. These distributions are checked for all levels of the telescope – for channel (single optical module (OM)), for section (section is a minimal structure unit of the detector, twelve linked optical modules), and for the whole cluster. The calibration systems used for PMT calibration, such as LED matrices or laser sources, operate with fixed frequency and emit the light that should worsen the shape of distributions, and therefore the fit quality decreases. On the other hand, unstable environmental conditions of the telescope~\cite{luminescence}, namely, a possible dynamical changing of the count rate during the seance of data taken (so-called "run") also lead to deterioration of distributions.

\subsection{Exponential distribution test}
\begin{figure*}
	\centering
	\begin{minipage}{16pc}
		\includegraphics[width=14pc] {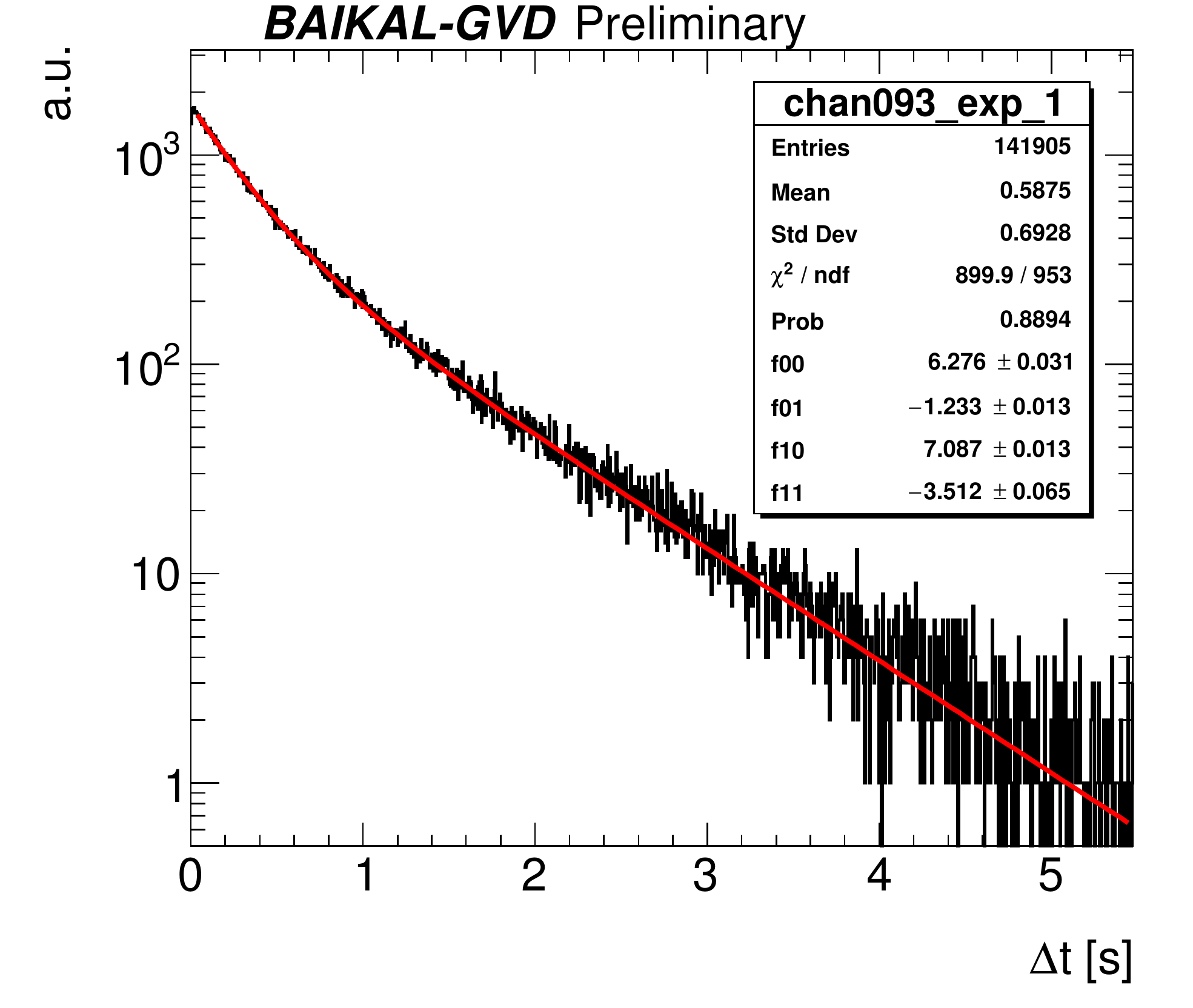}
	\end{minipage}
	\hspace{1pc}
	\qquad
	\begin{minipage}{16pc}
		\includegraphics[width=14pc] {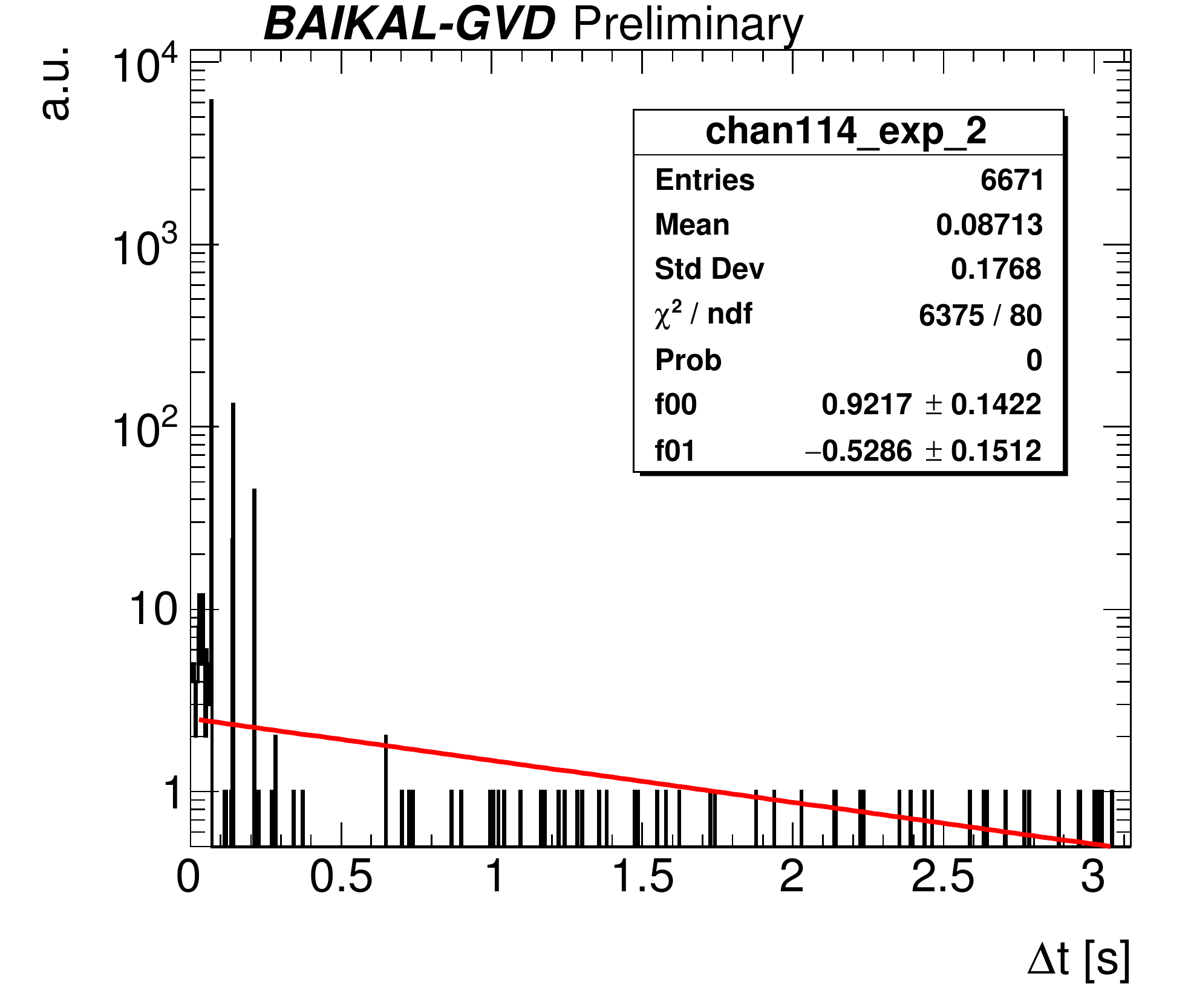}
	\end{minipage} 
	\caption{\label{fig_Exp} Exponential distributions for some selected channels: standard run (left), calibration run with laser source (right).}
\end{figure*}

The time difference between two consecutive events is described by the exponential function. Figure~\ref{fig_Exp} presents distributions for such time differences obtained for some selected channels. We take into account dynamical changing of the count rate during run, as it is shown in left panel. For such cases we fit the distribution by a set of exponential functions with different slopes, that correspond to different stable values of the count rate. There are clearly effect of laser source applied during calibration run, consequently the exponential distribution worsens and $\chi^2/NDF$ parameter is rather high (Figure~\ref{fig_Exp} (right)).

\subsection{Uniform distribution test}
\begin{figure*}
	\centering
	\begin{minipage}{16pc}
		\includegraphics[width=14pc] {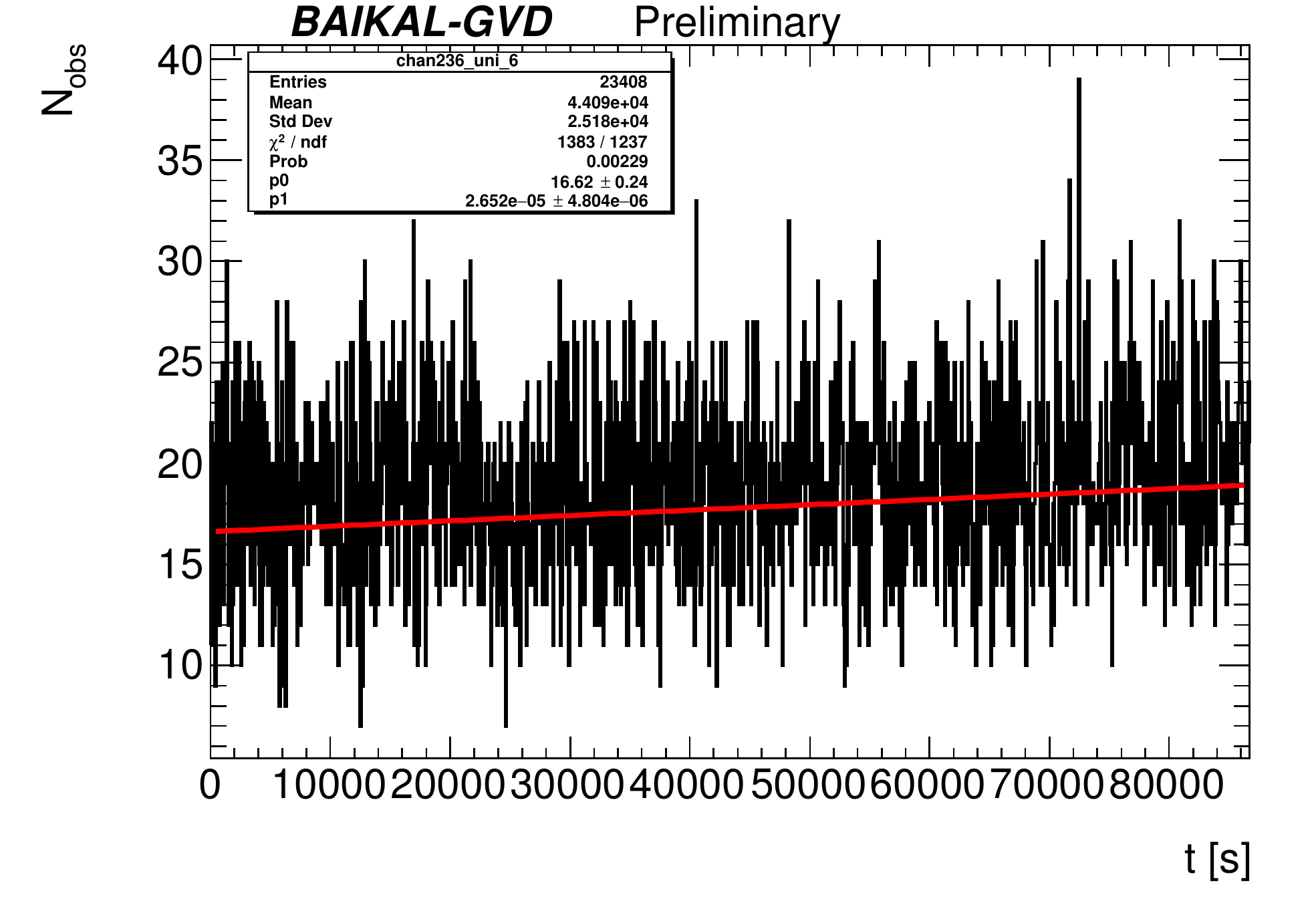}
	\end{minipage}
	\hspace{3mm}
	\qquad
	\begin{minipage}{16pc}
		\includegraphics[width=14pc] {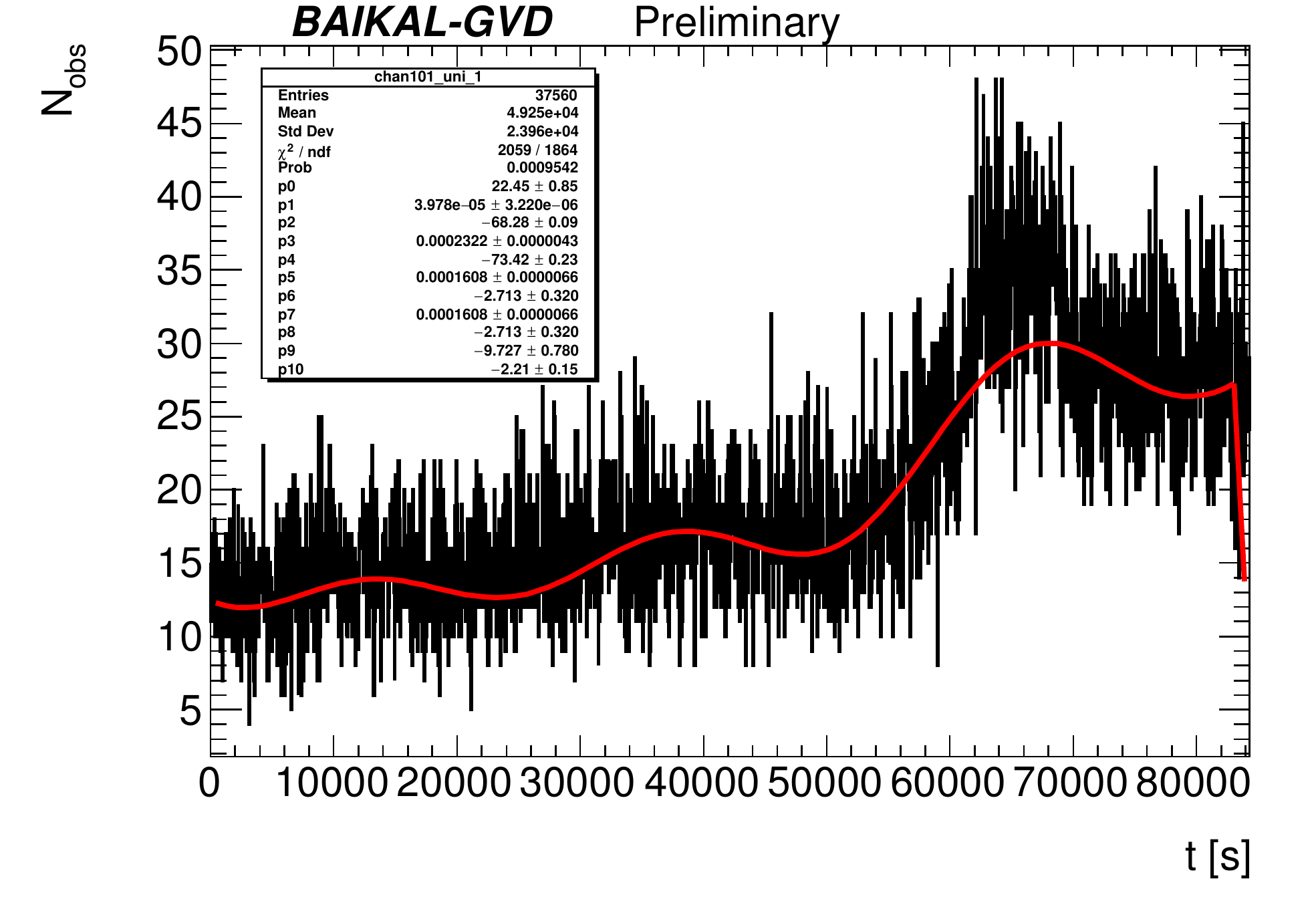}
	\end{minipage} 
	\caption{\label{fig_Uni} Uniform distributions for some selected channels: stable (left), and unstable (right) environmental condition.}
\end{figure*}

Uniform distribution is described the count rate of the recorded events during a run. It is expected that the distribution of the rate of atmospheric muons and lake random noise in stable environmental condition is linear with certain slope and should be described by the first order polynomial function as it is shown in Figure~\ref{fig_Uni} (left), where uniform distribution for some selected channel is shown. On the other hand, when environment during the run becomes unstable and the count rate quickly changes, we describe uniform distribution by a set of harmonic functions (see right panel in Figure~\ref{fig_Uni}).

\subsection{Poissonian distribution test}
\begin{figure*}
	\centering
	\begin{minipage}{16pc}
		\includegraphics[width=14pc] {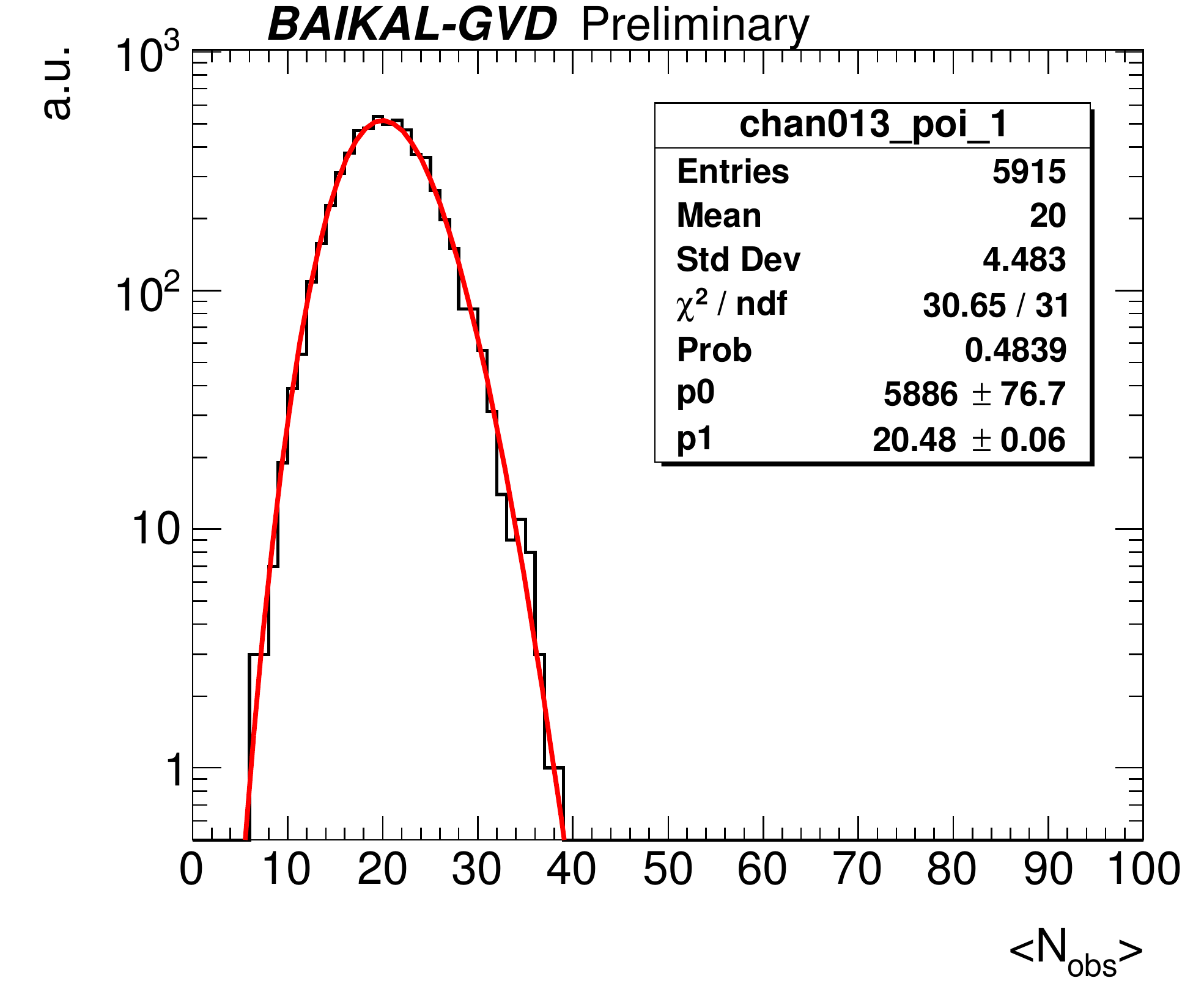}
	\end{minipage}
	\hspace{3mm}
	\qquad
	\begin{minipage}{16pc}
		\includegraphics[width=14pc] {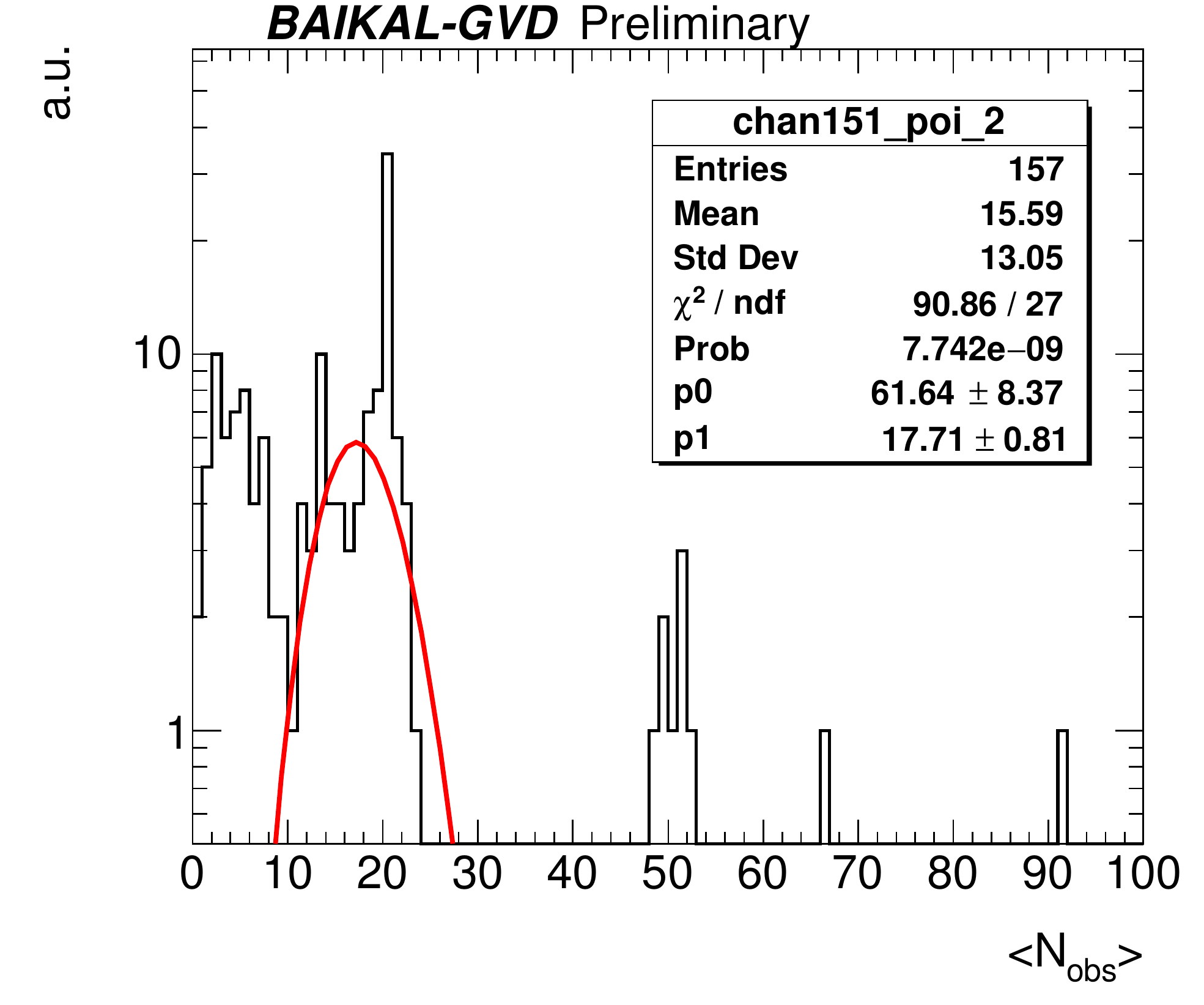}
	\end{minipage} 
	\caption{\label{fig_Poi} Poissonian distributions for some selected channels: standard run (left), calibration run with laser source (right).}
\end{figure*}

The expected number of the recorded events should follow the Poissonian distribution for any fixed time interval (Figure~\ref{fig_Poi}). In order to decrease an impact of the unstable environment, we divide the whole run to parts with stable count rate. For each of run parts the fixed time interval is chosen to have approximately 20 recorded events on average. As one can see in right panel of Figure~\ref{fig_Poi}, there is clear effect of the calibration run with the laser source switched on, when the distribution differs from poissonian shape, and the bin content in statistically suppressed regions of the distribution is high due to laser events, as well as the mean value of distribution differs from expected number of recorded events.

\section{Charge distributions test}
The second group of parameters which monitors the charge characteristics of signals from every PMTs is referred to charge distributions. The quality of the charge related data is estimated only for channel level. The trigger condition of the detector is fulfilled if neighbor pair of channels produces signals exceeding low charge threshold ($\sim$~1.5 p.e.) and high one ($\sim$~4 p.e.). In order to analyze trigger signals separately from non-trigger signals, a special algorithm is applied to raw data waveforms which reconstructs the trigger implementation for each individual event.

\subsection{Non-trigger signals}
\begin{figure} [h]
	\centering
	\includegraphics[width=16pc] {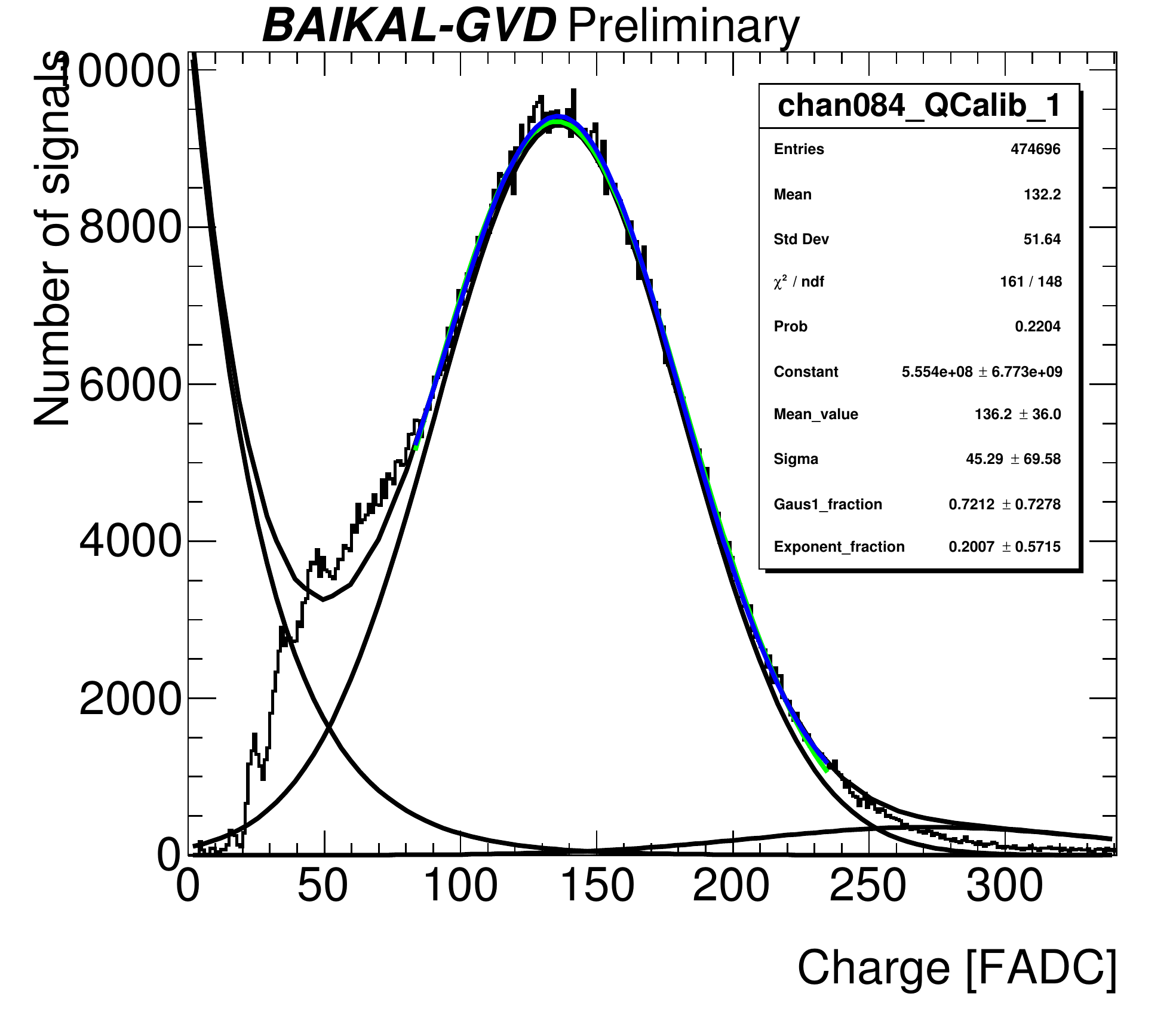}
	\caption{\label{fig_1PE} The charge distribution of non-trigger signals for a some selected channel.} 
\end{figure}

Any PMT signal exceeding the so-called "filter threshold" ($\sim$~0.5 p.e.) is recorded. Using information from channels that did not fire the trigger for a given event we obtain the combined charge distribution of non-trigger signals (Figure~\ref{fig_1PE}). Combined distribution consists of dominated single p.e. gaussian distribution and unessential contributions from dark currents of PMT (to the left of 1 p.e. peak), and 2 p.e. distribution (to the right). After fitting the combined distribution by a sum of one exponential function and two gaussian functions we can determine the position of the single p.e. peak for given channel in the current run, produced mainly by random noise. This value is used as calibration factor for recorded charge from FADC counts to p.e. values.

\begin{figure*}[h]
	\centering
	\begin{minipage}{16pc}
		\includegraphics[width=17pc] {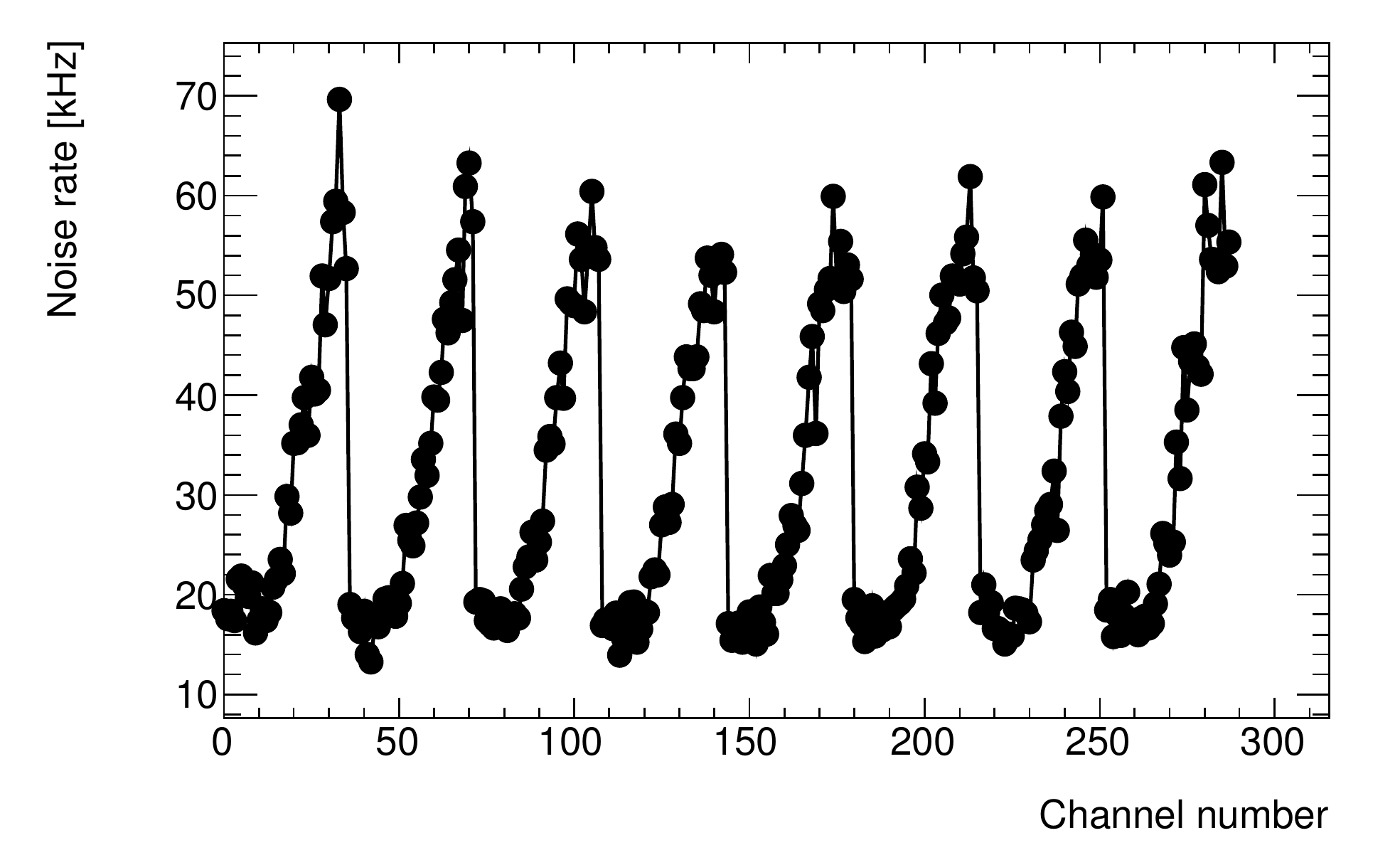}
	\end{minipage}
	\hspace{3mm}
	\qquad
	\begin{minipage}{16pc}
		\includegraphics[width=17pc] {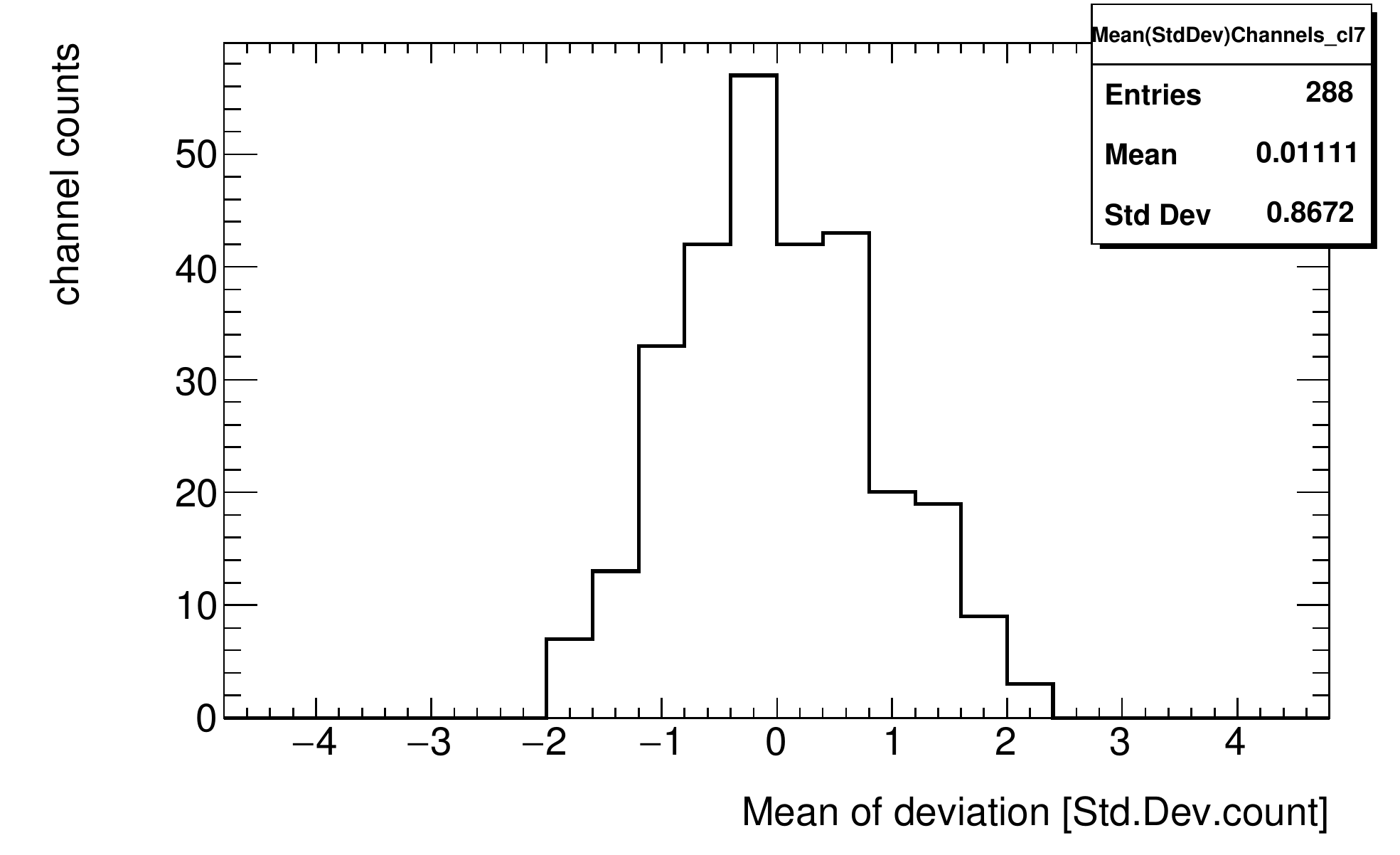}
	\end{minipage} 
	\caption{\label{fig_Noise} The noise rates for channels in some selected run (left). Channels noise deviation from the depth-averaged value (expressed in standard deviation counts) during over a half of season (right).}
\end{figure*}

Non-trigger signals also give the opportunity to estimate the noise count rate. Figure~\ref{fig_Noise} shows background noise characteristics of the collected data. It is expected that noise rate values should increase towards the higher vertical position of the channel on the string due to chemi luminescence (left panel)~\cite{luminescence}. We found that at every depth the channel noise rate is quite stable and practically do not depend on individual PMT, as it can be seen from the right panel of Figure~\ref{fig_Noise},  where the distribution of the mean noise rate deviations of each channel from depth-averaged rate obtained within over a half of season is shown.

\subsection{Trigger signals}
\begin{figure*} [h]
	\centering
	\begin{minipage}{16pc}
		\includegraphics[width=14pc] {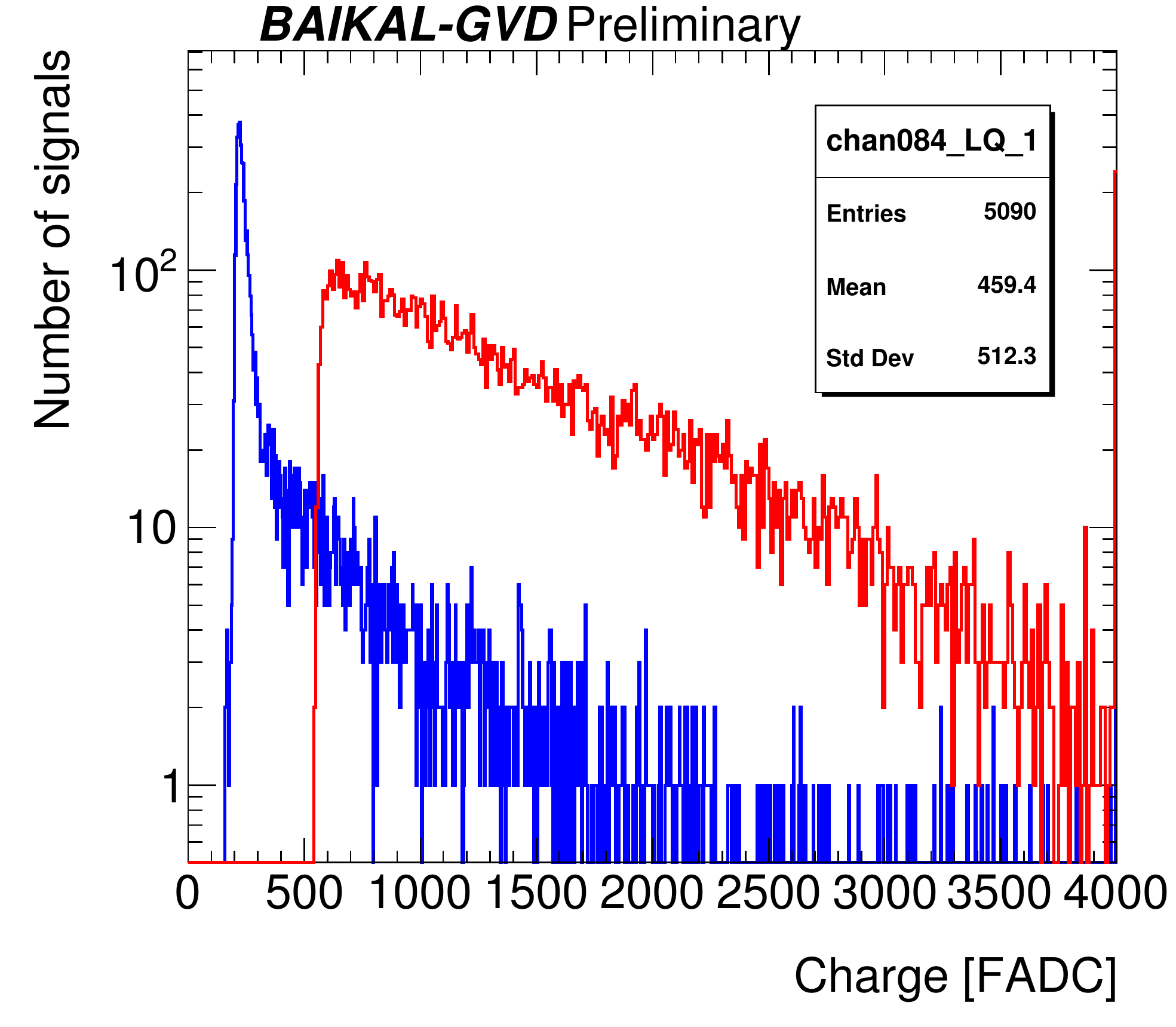}
	\end{minipage}
	\hspace{3mm}
	\qquad
	\begin{minipage}{16pc}
		\includegraphics[width=17pc] {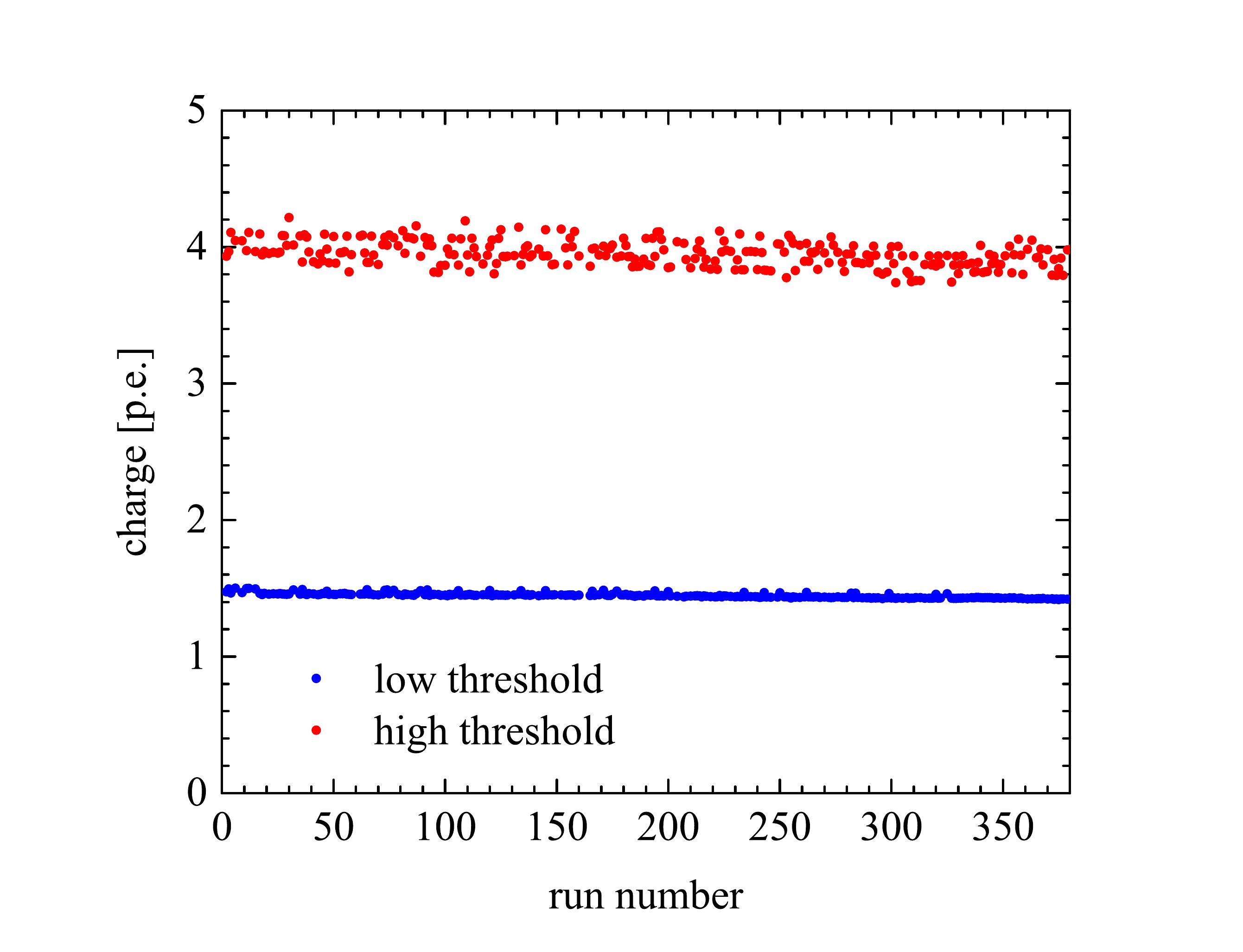}
	\end{minipage} 
	\caption{\label{fig_Thresholds} The signal charge values for some selected channel participating in the trigger (left). Low and high thresholds within over a half of year for some channel (right). Blue color stands for low threshold, and red color stands for high threshold.}
\end{figure*}

By using data on trigger signals, the DQM system performs monitoring of the measurements of the trigger charge thresholds. Low and high threshold values are set (or initialized) in the telescope configuration and are equal to $\sim$~1.5 p.e. and $\sim$~4 p.e., respectively. Initialized thresholds are set in 4 consecutive FADC counts, and it leads to an increase of the thresholds. We calculate the correction factor (that is equal to $\sim$~1.25) and then obtain a precise initialized threshold values in photoelectrons. Figure~\ref{fig_Thresholds} (left panel) shows the trigger thresholds that are measured in every run. The stability of these values is permanently monitored by the system for all channels during the run as well as the season.

\begin{figure*} [h]
	\centering
	\begin{minipage}{16pc}
		\includegraphics[width=17pc] {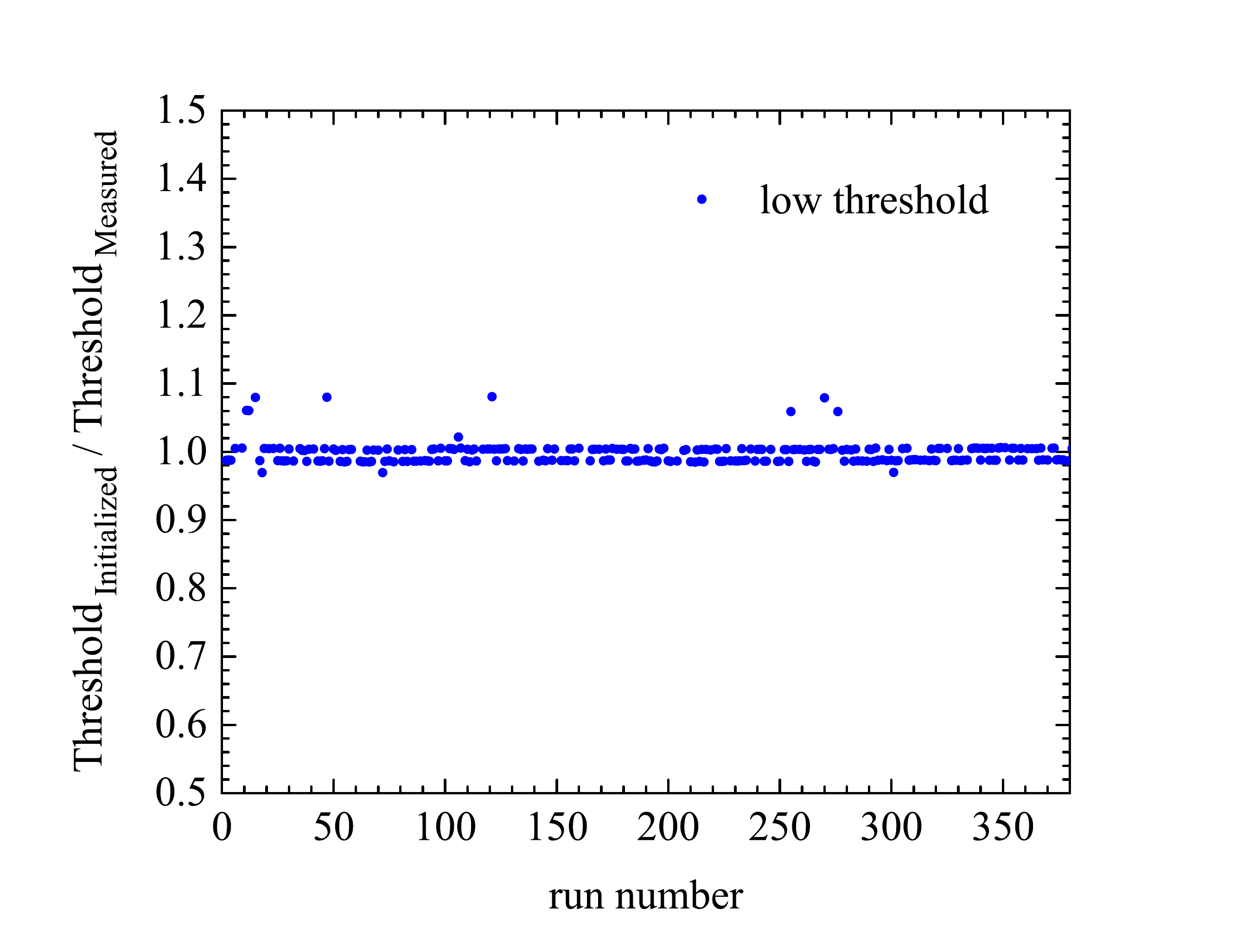}
	\end{minipage}
	\hspace{3mm}
	\qquad
	\begin{minipage}{16pc}
		\includegraphics[width=17pc] {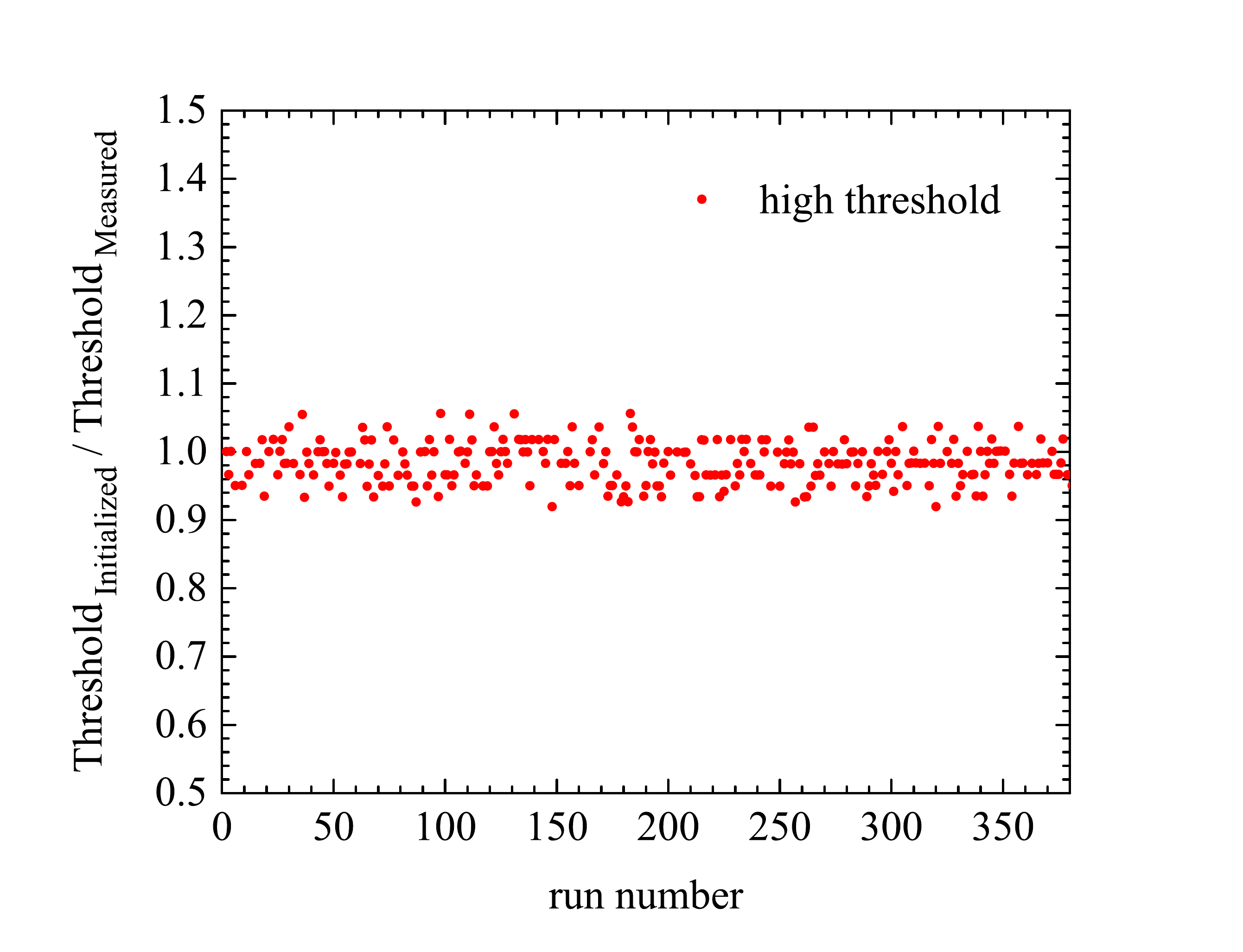}
	\end{minipage} 
	\caption{\label{fig_InitThresholds} Ratios of the initialized to measured trigger thresholds collected within over half of a year for some selected channel. Left panel stands for low threshold, and right one stands for high threshold.}
\end{figure*}

We found that the low and high thresholds expressed in terms of p.e. are quite stable for channels within season as it is shown in Figure~\ref{fig_Thresholds} (right), where threshold measurements over half of a year for some selected channel are plotted. Also, DQM system controls a deviation of measured thresholds from initialized values. Figure~\ref{fig_InitThresholds} shows the ratios of initialized to measured thresholds for some selected channel, which were collected during half of a year. We found a good agreement between initialized and measured thresholds within large data sample. As expected, these values are close to 1 and maximal deviations are smaller than 10\%.

\section{Quality estimation algorithm}
The purpose of the quality estimation algorithm is to obtain the quality status of the recorded data for each telescope unit. Channel, section and cluster levels are considered independently. Fit quality is estimated via $\chi^2/NDF$ parameter with threshold values for good (<2), normal (<4), and bad (>4) data. Fit quality estimations for each distribution are summed up as logical \textit{and}. There are eleven ranking markers for data quality status (codes):

~\

0 – excluded by configuration (defined by detector performance experts)

1 – empty data (<100 events in channel)

2 – good data

3 – normal data

4 – bad data

5 – good data but periodic light source is detected

6 – normal data but periodic light source is detected

7 – bad data but periodic light source is detected

8 – good data but unstable environmental conditions are detected

9 – normal data but unstable environmental conditions are detected

10 – bad data but unstable environmental conditions are detected

\section{Graphical user interface}
\begin{figure} [h]
	\centering
	\includegraphics[width=25pc] {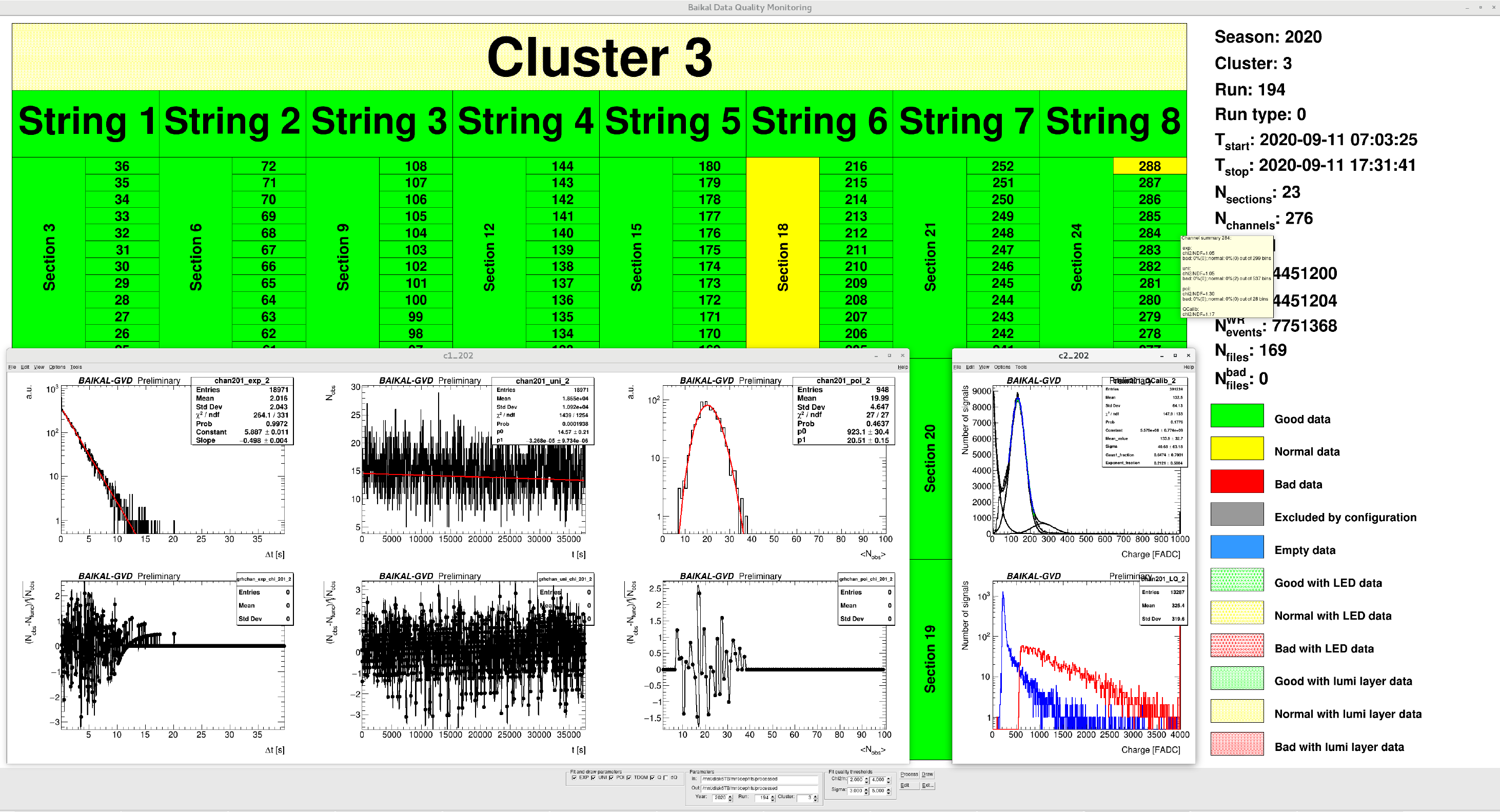}
	\caption{\label{fig_GUI} Graphical user interface of the DQM system.} 
\end{figure}

The DQM system is integrated within our unified software framework "BARS" and operates automatically. The results obtained by system are placed into the central database, and are showed on the dashboard web service. Also, the DQM system has a graphical user interface, that allows use the DQM outputs in off-line mode (see Figure~\ref{fig_GUI}). There is opportunity to check quality estimations for selected parameters (for example for exponential distribution, or for exponential distribution with poissonian one and so on), as well as use combined result obtained by using all distributions. An explicit view of the distributions with fit function can be shown along with graphs which represent the deviation of distribution from the obtained fit-function for all bins. Also, by hovering the mouse one can see short statistic box with fit quality results. Any level of the telescope can be seen independently and supporting information like timestamps of start and stop run, total number of events in run, and so on is presented.

\section{Summary}
We present the system which using to provide the quality estimations of recorded data of the Baikal-GVD. Data quality monitoring system operates automatically within Baikal-GVD’s unified software framework "BARS" and it’s graphical user interface allows to analyze the data quality in an off-line mode. DQM system takes into account a possible quick change of the telescope's background environment. By means of some examples of the data recorded during the 2020 season it was shown that applying various parameters for data quality analysis allows to estimate the quality of obtained data efficiently.

\section{Acknowledgements}
The work was partially supported by RFBR grant 20-02-00400. The CTU group acknowledges the support by European Regional Development Fund-Project No. CZ.02.1.01/0.0/0.0/16\_019/0000766. We also acknowledge the technical support of JINR staff for the computing facilities (JINR cloud).

%
%
%


\begin{thebibliography}{99}

\bibitem{GVD1}
A.V.~Avrorin et al., Baikal-GVD coll., Baikal-GVD Experiment. Phys. Atom. Nucl. {\bf 83}, 916 (2020)

\bibitem{GVD2}
I.A.~Belolaptikov et al., Baikal-GVD coll., Neutrino Telescope in Lake Baikal: Present and Nearest Future. {\em PoS} {\bfseries ICRC2021} (these proceedings) 002

\bibitem{BARS}
B.A.~Shaybonov et al., Baikal-GVD coll., Automatic data processing for Baikal-GVD neutrino observatory. {\em PoS} {\bfseries ICRC2021} (these proceedings) 1040

\bibitem{luminescence}
R.~Dvornick\'{y} et al., Baikal-GVD coll., The Baikal-GVD neutrino telescope as an instrument for studying Baikal water luminescence. {\em PoS} {\bfseries ICRC2021} (these proceedings) 1113


\end{thebibliography}
\end{document}